\documentclass[conference]{IEEEtran}
\IEEEoverridecommandlockouts
\renewcommand{\thesection}{\Roman{section}}
\usepackage{booktabs}
\usepackage{algorithm}
\usepackage[usenames, dvipsnames]{xcolor}
\usepackage{algorithmic}
\usepackage{colortbl}
\usepackage{caption}
\usepackage{graphicx}
\usepackage{array}
\usepackage{adjustbox}
\usepackage{cite}
\usepackage{multirow}
\usepackage{amsmath,amssymb,amsfonts}
\usepackage{textcomp}
\usepackage{comment}
\usepackage{balance}
\usepackage{fancyhdr}
\usepackage{hyperref}
\hypersetup{
    colorlinks=true,
    linkcolor=blue,
    citecolor=blue,
    urlcolor=blue,
    pdftitle={PIE-ADA: Physics-Informed Ensemble with Adaptive Data Augmentation for Photometric Transient Classification},
    pdfauthor={Deba Priyo Guha, Fariya Tabassum}
}

\def\BibTeX{{\rm B\kern-.05em{\sc i\kern-.025em b}\kern-.08em
    T\kern-.1667em\lower.7ex\hbox{E}\kern-.125emX}}

\renewcommand{\thanks}[1]{\unskip}

\fancypagestyle{ieeecopyright}{%
  \fancyhf{}%
  \fancyfoot[C]{\scriptsize\textit{This is the author accepted manuscript of a paper published at the 2026 IEEE 2nd International Conference on Quantum Photonics, Artificial Intelligence, and Networking (QPAIN 2026). Published version: \href{https://doi.org/10.1109/QPAIN69676.2026.11546175}{doi:10.1109/QPAIN69676.2026.11546175}. \textcopyright~2026 IEEE. Personal use of this material is permitted. Permission from IEEE must be obtained for all other uses, in any current or future media, including reprinting/republishing this material for advertising or promotional purposes, creating new collective works, for resale or redistribution to servers or lists, or reuse of any copyrighted component of this work in other works.}}%
}

\begin{document}

\title{PIE-ADA: Physics-Informed Ensemble with Adaptive Data Augmentation for Photometric Transient Classification}

\author{
\IEEEauthorblockN{Deba Priyo Guha\thanks{Center for Astronomy, Space Science and Astrophysics, Independent University, Bangladesh, Dhaka 1229, Bangladesh. Email: debapriyoguha@gmail.com.}\textsuperscript{1,2},
Fariya Tabassum\thanks{Department of Electrical and Computer Engineering, Rajshahi University of Engineering and Technology, Kazla, Rajshahi 6204, Bangladesh. Email: fariya@ece.ruet.ac.bd.}\textsuperscript{2}}
\IEEEauthorblockA{\textsuperscript{1}Center for Astronomy, Space Science \& Astrophysics,\\
Independent University, Bangladesh, Dhaka 1229, Bangladesh\\
\textsuperscript{2}Department of Electrical \& Computer Engineering,\\
Rajshahi University of Engineering \& Technology, Kazla, Rajshahi 6204, Bangladesh\\
\textnormal{\textsuperscript{1}debapriyoguha@gmail.com,\ \textsuperscript{2}fariya@ece.ruet.ac.bd}}
}

\maketitle

\begin{abstract}
The upcoming Large Synoptic Survey Telescope (LSST) is expected to observe approximately 10 million astronomical transient events per night, creating an urgent need for automated classification. A key challenge is the extreme class imbalance in transient datasets, where rare event types represent less than 1\% of all observations. This paper presents PIE-ADA (Physics-Informed Ensemble with Adaptive Data Augmentation), a framework that generates physically realistic synthetic light curves for underrepresented classes using astrophysically motivated transformations. PIE-ADA applies four augmentation operations, namely correlated noise injection, cosmological time dilation, wavelength-dependent dust extinction, and observation phase shifting, while enforcing physical constraints to prevent unrealistic samples. We extract 271 multi-scale features from six photometric passbands covering statistical, temporal, peak, color, and frequency-domain properties. Evaluated on the PLAsTiCC dataset (7,848 original objects augmented to 8,148 across 14 classes), five classifiers were compared using stratified 5-fold cross-validation. LightGBM achieved the best performance with a weighted log loss of 0.5763 ($\pm$0.0083) and 80.33\% accuracy, improving over Random Forest, Extra Trees, and Neural Network baselines by 24--49\% in log loss. The framework is computationally efficient, completing the full pipeline in under 37 minutes and classifying individual objects in less than 0.05 seconds, making it suitable for real-time LSST alert processing.
\end{abstract}


\section{Introduction}
\label{sec:introduction}
\IEEEPARstart{T}{he} next generation of astronomical surveys will fundamentally change how we observe the transient sky. The Vera C. Rubin Observatory's Legacy Survey of Space and Time (LSST), scheduled to begin operations in 2025, will produce roughly 10 million transient alerts every night \cite{Ivezić_2019}. This volume represents a thousandfold increase compared to current capabilities, offering remarkable scientific potential alongside serious computational demands.

Astronomical transients, objects that change in brightness over time, include phenomena such as supernovae, kilonovae, tidal disruption events, active galactic nuclei, and microlensing events. Rapid and accurate classification of these events is essential for multi-messenger astronomy, as it allows researchers to prioritize spectroscopic follow-up observations before targets fade \cite{Narayan_2018}. Since spectroscopic confirmation is both expensive and time-consuming, reliable photometric classification from multi-band light curves becomes critical.

\subsection{The Class Imbalance Challenge}

A major difficulty in transient classification is the severe imbalance among event types. Common transients such as Type Ia supernovae can dominate datasets, while scientifically valuable rare events like superluminous supernovae or kilonovae may account for less than 0.1\% of all samples \cite{10.1093/mnras/stz2362}. Standard machine learning models trained on such skewed distributions tend to favor majority classes, resulting in poor recall for rare but important event types.

The PLAsTiCC challenge \cite{plasticc2018} illustrated this problem clearly by providing a realistic simulated dataset containing 14 transient classes with imbalance ratios ranging from 0.3\% to 30\%. Top-performing solutions in that competition employed Gaussian process regression \cite{Boone_2019}, ensemble stacking, and extensive feature engineering. However, many leading approaches sacrificed interpretability for marginal performance gains and required substantial computational resources.

\subsection{Our Contributions}

We propose PIE-ADA (Physics-Informed Ensemble with Adaptive Data Augmentation), a classification framework that tackles class imbalance through physics-guided synthetic data generation. The term ``adaptive'' refers to the framework's ability to automatically identify underrepresented classes and adjust the augmentation intensity based on the degree of imbalance, generating more synthetic samples for severely rare classes while leaving well-represented classes unchanged. Our main contributions are:

\begin{itemize}
    \item \textbf{Physics-informed augmentation:} Synthetic light curves for rare classes are generated using physically realistic transformations including time dilation, wavelength-dependent extinction, correlated noise injection, and phase shifting, preserving temporal and spectral properties that generic methods like SMOTE would violate.

    \item \textbf{Multi-scale feature extraction:} A total of 271 interpretable features are computed across statistical, temporal, peak, color, and frequency-domain categories from six photometric passbands.
    
    \item \textbf{Systematic model comparison:} Five algorithms are benchmarked under identical conditions, with LightGBM achieving the lowest weighted log loss of 0.5763 and 80.33\% accuracy, outperforming baselines by 24--49\%.
\end{itemize}

\section{Literature Review}
\label{sec:related}
\subsection{Transient Classification Methods}

Early approaches to transient classification relied on template fitting and color-magnitude diagrams. Machine learning methods brought significant improvements: Lochner et al. \cite{Lochner_2016} showed that classifiers trained on light curve features could achieve competitive photometric supernova classification. Recurrent neural networks were later applied by Charnock and Maund \cite{Charnock_2017}, while Muthukrishna et al. \cite{Muthukrishna_2019} developed RAPID, an RNN-based system requiring careful hyperparameter tuning. M{\"o}ller and de Boissi{\`e}re \cite{Moller_2020} introduced SuperNNova, a Bayesian neural network providing uncertainty estimates alongside predictions.

The PLAsTiCC competition produced notable advances. Boone et al. \cite{Boone_2019} won first place (0.51 weighted log loss) using Gaussian Process regression combined with XGBoost. Villar et al. \cite{Villar_2019} demonstrated that ensemble approaches with physics-based features consistently outperformed single models. More recently, Qu and Sako \cite{Qu_2022} developed SCONE, which uses two-dimensional Gaussian processes to create flux heatmaps for convolutional neural network classification. Burhanudin and Maund \cite{Burhanudin_2022} achieved an AUC of 0.945 on PLAsTiCC data through transfer learning across surveys.

\subsection{Handling Class Imbalance}

Class imbalance is widespread in transient datasets where rare events may constitute less than 1\% of samples \cite{10.1093/mnras/stz2362}. SMOTE \cite{Chawla_2002} addresses this by interpolating between minority class neighbors in feature space, but such generic approaches ignore domain-specific constraints and can produce physically unrealistic samples. Leoni et al. \cite{Leoni_2022} showed that active learning can reduce required training data by 40\%.

For time-series data, augmentation strategies include jittering, scaling, time warping, and window slicing \cite{Wen_2021}. Our work differs by performing augmentation directly on raw light curves using physically motivated transformations, preserving both temporal correlations and spectral energy distributions that abstract feature-space methods would disrupt.

\section{Methodology}
\label{sec:methodology}
\begin{figure}[h]
\centering
\includegraphics[width=0.48\textwidth]{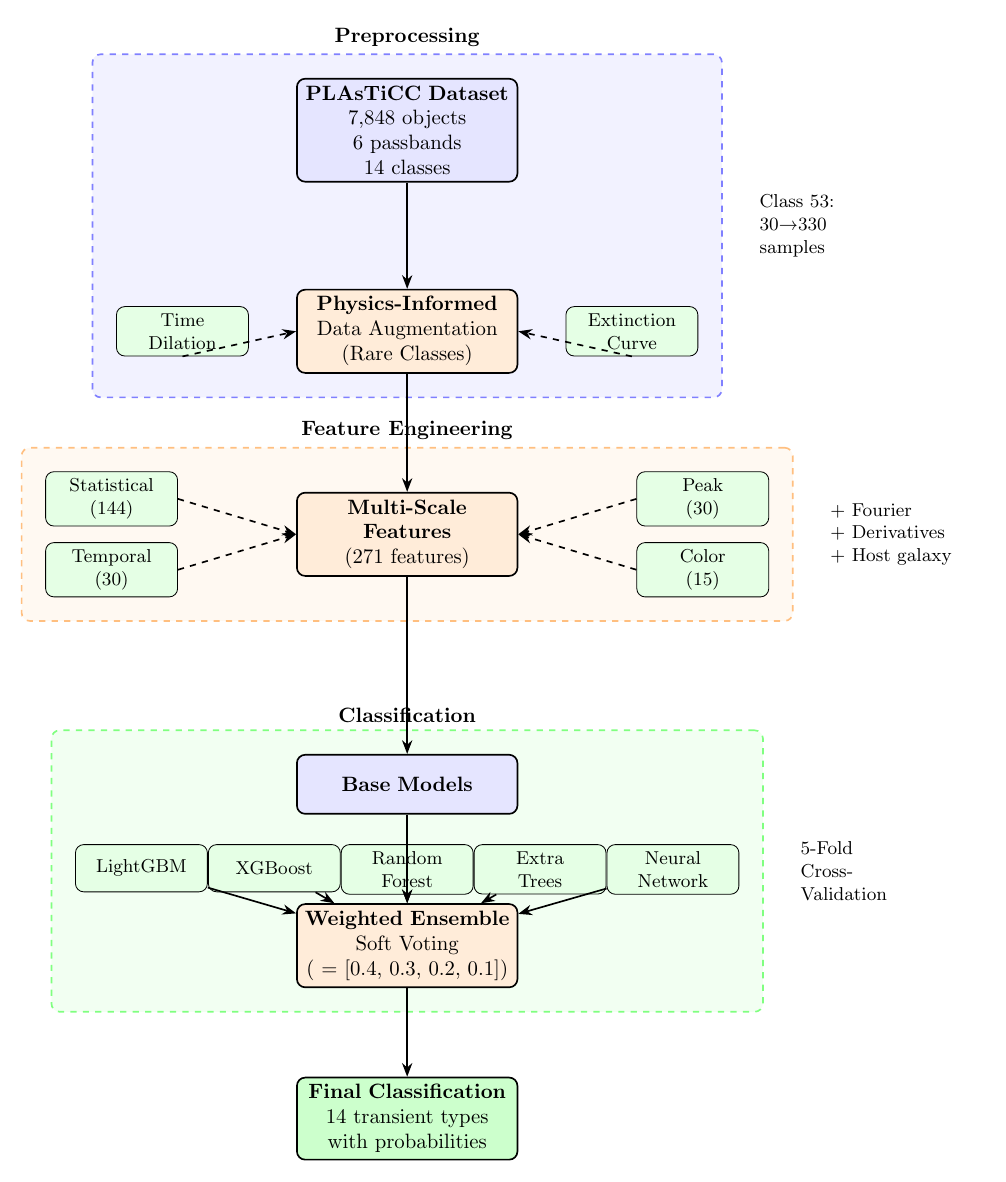}
\caption{PIE-ADA pipeline: physics-informed augmentation generates synthetic samples for rare classes, 271 multi-scale features are extracted, and five classifiers are compared with optional ensemble combination.}
\label{fig:pipeline}
\end{figure}

Our methodology consists of four components: (1) physics-informed data augmentation, (2) multi-scale feature engineering, (3) model training with weighted loss, and (4) ensemble construction. Figure~\ref{fig:pipeline} illustrates the complete PIE-ADA pipeline.

\subsection{Dataset}

We use the PLAsTiCC training set \cite{plasticc2018}, which contains 7,848 astronomical transients observed across six photometric passbands ($u, g, r, i, z, y$) with 1,421,705 individual photometric measurements. Each object belongs to one of 14 classes exhibiting severe imbalance: Type Ia supernovae (class 90) comprise 29.5\% of the dataset, while M-dwarf flares (class 53) represent only 0.4\% with just 30 objects. Table~\ref{tab:class_dist} presents the class distribution before and after augmentation.

\begin{table}[h]
\centering
\caption{Class Distribution Before and After Augmentation}
\label{tab:class_dist}
\small
\begin{tabular}{clcc}
\toprule
\textbf{ID} & \textbf{Class} & \textbf{Original} & \textbf{Final} \\
\midrule
90 & SN Ia & 2313 & 2313 \\
42 & SN II & 1193 & 1193 \\
65 & Eclipsing Binary & 981 & 981 \\
16 & Mira Variable & 924 & 924 \\
15 & TDE & 495 & 495 \\
62 & SN Iax & 484 & 484 \\
88 & AGN & 370 & 370 \\
92 & RR Lyrae & 239 & 239 \\
67 & SN 91bg & 208 & 208 \\
52 & SN Ibc & 183 & 183 \\
95 & SLSN-I & 175 & 175 \\
6 & $\mu$-Lens Single & 151 & 151 \\
64 & Kilonova & 102 & 102 \\
53 & M-dwarf Flare & 30 & 330 \\
\midrule
\multicolumn{2}{c}{\textbf{Total}} & \textbf{7,848} & \textbf{8,148} \\
\bottomrule
\end{tabular}
\end{table}

\subsection{Physics-Informed Data Augmentation}

Standard augmentation techniques such as SMOTE generate synthetic samples through feature-space interpolation, which can produce physically unrealistic light curves that violate astrophysical constraints. PIE-ADA instead augments rare classes (those with fewer than 100 samples) directly in the observational domain using transformations grounded in physical phenomena.

The adaptive component of PIE-ADA operates as follows: the framework first scans the training set to identify classes below a configurable sample threshold (set to 100 in our experiments). For each identified rare class, the augmentation multiplier is determined by the ratio between the threshold and the actual class count, ensuring that more severely underrepresented classes receive proportionally greater augmentation. This automatic adjustment removes the need for manual tuning of per-class augmentation parameters.

\subsubsection{Augmentation Transformations}

For each rare-class object, 10 synthetic variants are generated by randomly sampling from four operations:

\textbf{1) Correlated Noise Injection:} Photometric measurements exhibit temporal correlations from atmospheric and instrumental effects. We model this using a first-order autoregressive process:
\begin{equation}
n_t = \alpha n_{t-1} + (1-\alpha)\epsilon_t
\end{equation}
where $\epsilon_t \sim \mathcal{N}(0, \sigma^2)$, $\alpha = 0.6$ controls correlation strength, and noise is scaled by flux uncertainty: $\delta f_{\lambda,t} = 0.2 \times \sigma_{f,\lambda,t} \times n_t$.

\textbf{2) Time Dilation:} Cosmological redshift stretches observed timescales by $(1+z)$. We apply temporal scaling $\tau \sim \mathcal{U}(0.95, 1.05)$:
\begin{equation}
\text{MJD}'_t = \text{MJD}_0 + (t - t_0) \cdot \tau
\end{equation}

\textbf{3) Wavelength-Dependent Extinction:} Interstellar dust absorbs shorter wavelengths preferentially, following the Cardelli law \cite{1989ApJ...345..245C}:
\begin{equation}
f'_\lambda = f_\lambda \cdot 10^{-0.4 E(B\text{-}V) \cdot k_\lambda}
\end{equation}
with $E(B\text{-}V) \sim \mathcal{U}(0, 0.2)$ and passband coefficients $k_u{=}1.0$, $k_g{=}0.75$, $k_r{=}0.5$, $k_i{=}0.38$, $k_z{=}0.25$, $k_y{=}0.13$.

\textbf{4) Phase Shifts:} Observation epochs vary due to survey cadence. Light curves are shifted by $\Delta t \sim \mathcal{U}(-3, 3)$ days.

Physical constraints are enforced to prevent unrealistic outputs: flux values are bounded below by $f_\lambda > -0.2 \times f_{\lambda,\text{peak}}$, photometric redshifts are perturbed within $\pm 0.02$, and temporal ordering is preserved. Algorithm~\ref{alg:augmentation} details the complete procedure.

\begin{algorithm}[h]
\caption{Physics-Informed Data Augmentation}
\label{alg:augmentation}
\begin{algorithmic}[1]
\REQUIRE Light curve $L = \{(t_i, f_{\lambda,i}, \sigma_i)\}$, class $c$
\ENSURE Augmented set $\{L'_1, \ldots, L'_N\}$
\FOR{$k = 1$ to $N = 10$}
    \STATE $L' \leftarrow L$ \COMMENT{copy original}
    \STATE Sample $\tau \sim \mathcal{U}(0.95, 1.05)$, $E(B\text{-}V) \sim \mathcal{U}(0, 0.2)$, $\Delta t \sim \mathcal{U}(-3, 3)$
    \FOR{each passband $\lambda$}
        \STATE Generate noise: $n_t = 0.6 n_{t-1} + 0.4 \epsilon_t$
        \STATE $f'_{\lambda,t} \leftarrow f_{\lambda,t} \cdot 10^{-0.4 E(B\text{-}V) k_\lambda} + 0.2 \sigma_t n_t$
        \STATE $t' \leftarrow t_0 + (t - t_0)\tau + \Delta t$
    \ENDFOR
    \STATE Enforce $f' > -0.2 \max(f_\lambda)$
\ENDFOR
\RETURN $\{L'_1, \ldots, L'_N\}$
\end{algorithmic}
\end{algorithm}

This process generated 43,910 new photometric observations for class 53 (M-dwarf flares), increasing its sample count from 30 to 330 objects.

\subsection{Multi-Scale Feature Engineering}

We extract 271 features per object across six categories:

\textbf{Statistical (145):} Mean, standard deviation, skewness, kurtosis, MAD, percentiles, and signal-to-noise ratio per passband.

\textbf{Temporal (6):} Duration, flux variation rate, detection rate, and gap statistics.

\textbf{Peak (31):} Rise and fall rates, peak timing per passband.

\textbf{Color (16):} Magnitude differences between passband pairs: $C_{\lambda_1\text{-}\lambda_2} = -2.5 \log_{10}(\langle f_{\lambda_1} \rangle / \langle f_{\lambda_2} \rangle)$.

\textbf{Frequency-domain (71):} Fourier amplitudes, flux ratios, percentiles, and derivatives.

\textbf{Host galaxy (2):} Photometric redshift and distance modulus.

\subsection{Model Training and Ensemble}

Five algorithms are trained using weighted logarithmic loss consistent with PLAsTiCC evaluation: LightGBM (2000 trees, learning rate 0.01, 300 leaves), XGBoost (1000 trees), Random Forest (1000 trees, entropy criterion), Extra Trees (1000 trees), and a Neural Network (271-100-50-14 architecture). All models use stratified 5-fold cross-validation with fixed random seeds.

The ensemble combines predictions through soft voting weighted by validation performance: LightGBM (0.40), XGBoost (0.30), Random Forest (0.20), Extra Trees (0.10).


\section{Results and Discussions}
\label{sec:experiments}
\subsection{Augmentation Impact}

Physics-informed augmentation increased rare class samples by 10$\times$ while maintaining physical realism. M-dwarf flares (class 53) grew from 30 to 330 objects (43,910 new photometric observations). Visual inspection confirmed that augmented light curves preserve temporal structure and spectral characteristics of the original samples.

\subsection{Model Performance}

Table~\ref{tab:results} presents cross-validation and held-out test set results for all algorithms. LightGBM achieves the best cross-validation performance with a weighted log loss of 0.5763 ($\pm$0.0083) and 80.33\% accuracy. It outperforms XGBoost by 6.8\%, Random Forest by 31.9\%, Extra Trees by 45.0\%, and Neural Networks by 49.3\% in terms of log loss.

\begin{table}[h]
\centering
\caption{Model Performance Comparison (5-Fold CV and Test Set)}
\label{tab:results}
\begin{tabular}{lcccc}
\toprule
\textbf{Algorithm} & \textbf{CV Loss} & \textbf{CV Acc} & \textbf{Test Loss} & \textbf{Test Acc} \\
\midrule
LightGBM     & \textbf{0.576} & \textbf{0.803} & \textbf{0.590} & \textbf{0.801} \\
XGBoost      & 0.615 & 0.797 & 0.635 & 0.796 \\
Random Forest & 0.760 & 0.754 & 0.785 & 0.767 \\
Extra Trees   & 0.836 & 0.739 & 0.892 & 0.763 \\
Neural Network & 0.860 & 0.737 & 0.870 & 0.726 \\
Ensemble      & 0.624 & 0.799 & 0.643 & 0.795 \\
\bottomrule
\end{tabular}
\end{table}

\begin{figure}[h]
\centering
\includegraphics[width=0.48\textwidth]{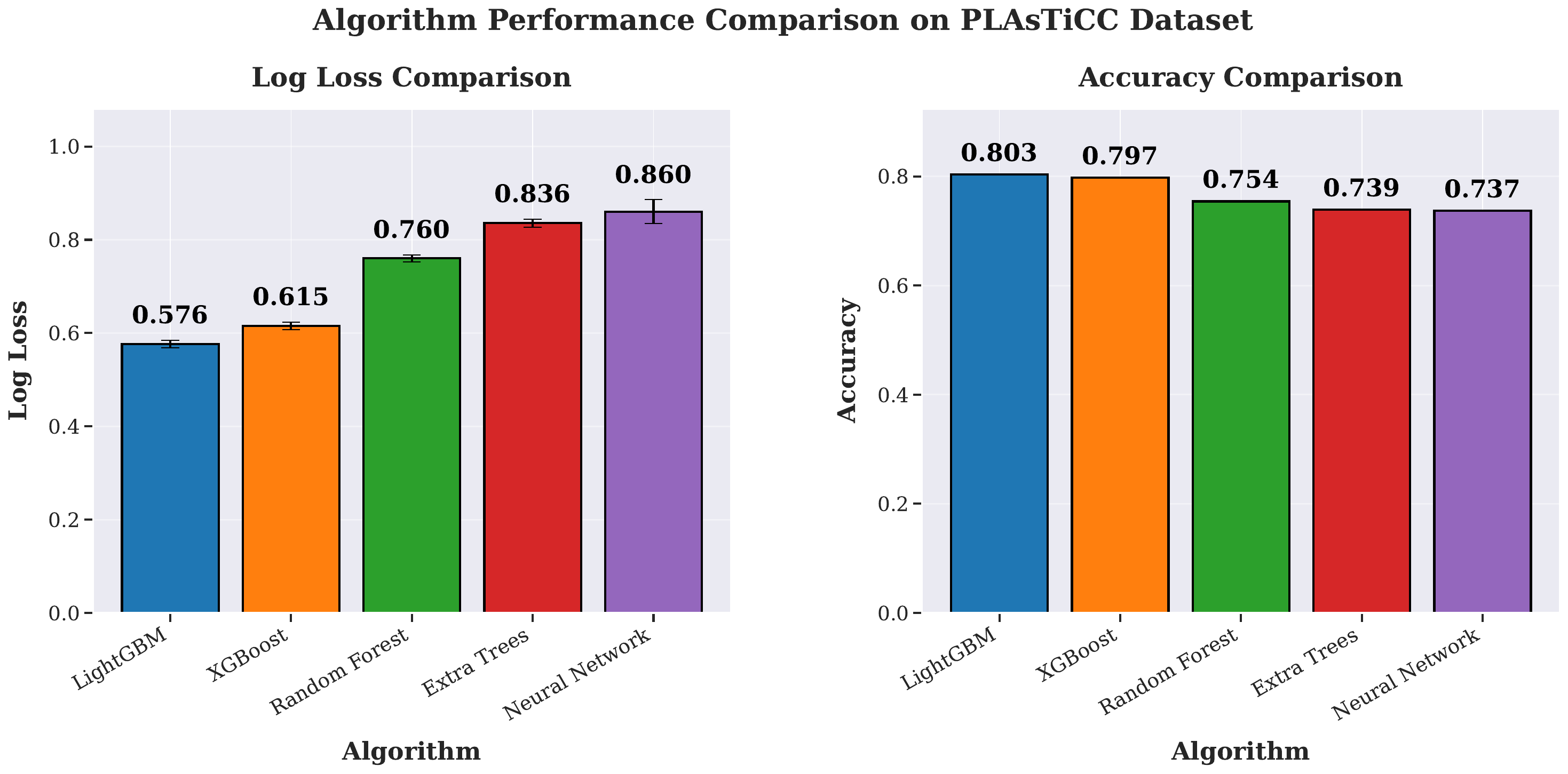}
\caption{Algorithm performance comparison showing weighted log loss (left) with standard deviation error bars and classification accuracy (right) across 5-fold cross-validation.}
\label{fig:algorithm_comparison}
\end{figure}

The standard deviations across folds are small for all gradient boosting methods (LightGBM: $\pm$0.008, XGBoost: $\pm$0.008), indicating stable performance. The Neural Network shows the highest variance ($\pm$0.026) due to sensitivity to random initialization. Figure~\ref{fig:algorithm_comparison} presents this comparison visually.

LightGBM performs well due to its leaf-wise tree growth strategy, efficient handling of high-dimensional sparse features, and built-in L1/L2 regularization that prevents overfitting on augmented samples.

\subsection{Feature Importance}

\begin{figure}[h]
\centering
\includegraphics[width=0.48\textwidth]{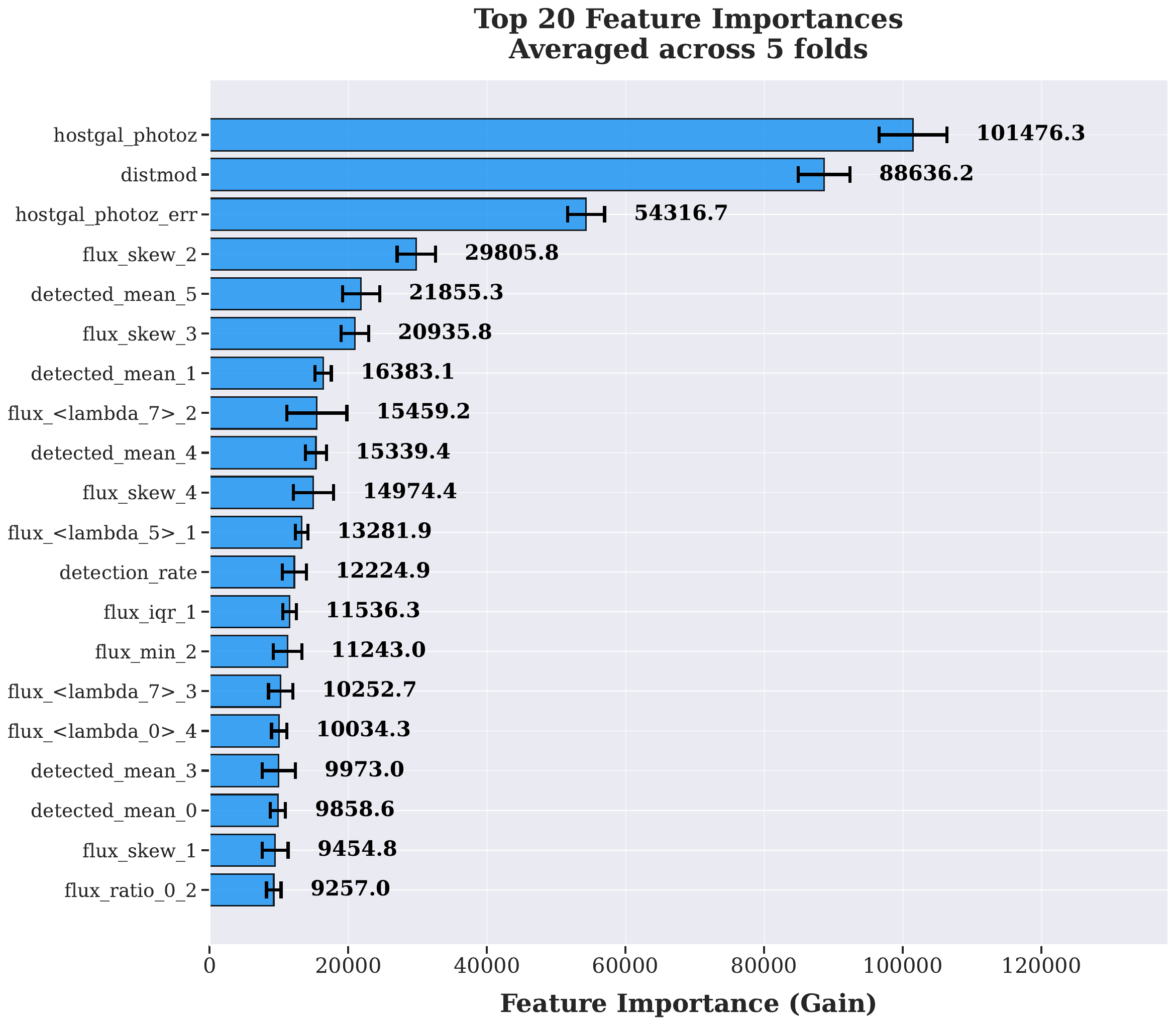}
\caption{Top 20 features by mean importance gain averaged across 5-fold cross-validation. Error bars indicate standard deviation across folds.}
\label{fig:feature_importance}
\end{figure}

Figure~\ref{fig:feature_importance} shows the top 20 features ranked by mean gain across folds. Host galaxy photometric redshift (101,476) and distance modulus (88,636) are the most influential, confirming that contextual metadata plays a central role in classification. Flux skewness across passbands and detection statistics further help separate temperature-sensitive classes. Fourier-domain features contribute relatively little, suggesting that periodic transients can be identified using simpler temporal statistics.

\subsection{Per-Class Analysis}

\begin{figure}[h]
\centering
\includegraphics[width=0.48\textwidth]{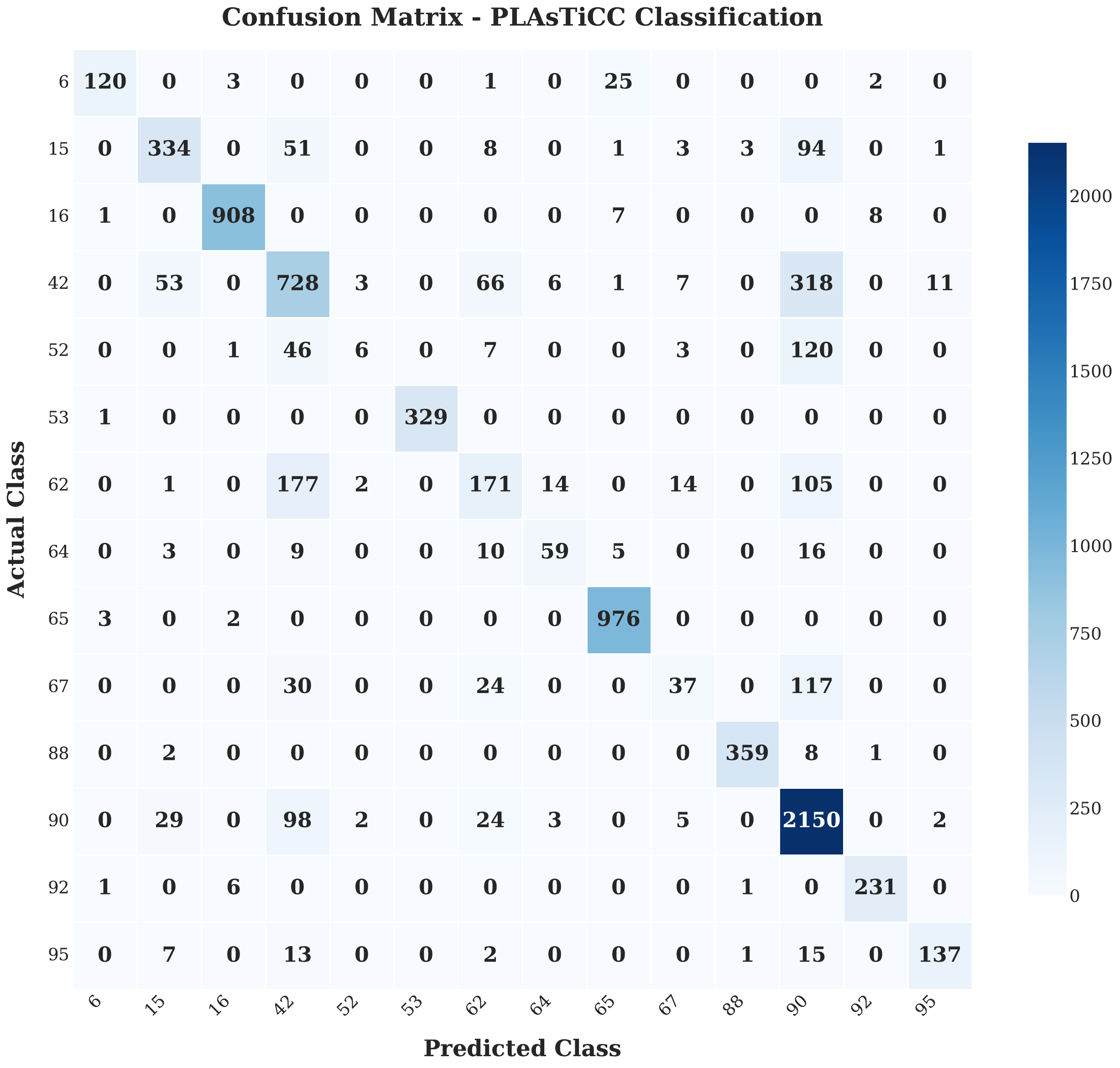}
\caption{Confusion matrix for LightGBM cross-validation predictions showing strong performance on most classes with expected confusion among spectroscopically similar supernova subtypes.}
\label{fig:confusion_matrix}
\end{figure}

Figure~\ref{fig:confusion_matrix} presents the confusion matrix. Classes with recall above 95\% include M-dwarf flares (class 53: 100\%), eclipsing binaries (class 65: 99\%), AGN (class 88: 97\%), and RR Lyrae (class 92: 97\%). The augmented class 53 achieves perfect classification, demonstrating the effectiveness of physics-informed augmentation.

Challenging classes with recall below 20\% include Type Ibc supernovae (class 52: 3\%) and Type 91bg supernovae (class 67: 18\%). These classes are frequently confused with Type Ia supernovae (class 90) because their photometric signatures overlap substantially, and distinguishing them typically requires spectroscopic observations of ejecta composition.


\subsection{Ensemble Analysis}

The soft voting ensemble (weighted combination of four models) achieves a log loss of 0.6239, which is 8.3\% worse than LightGBM alone. Weaker models (Extra Trees: 0.836, Neural Network: 0.860) degrade ensemble quality despite optimized weighting. For operational deployment, standalone LightGBM is recommended as it provides both the best accuracy and lowest computational overhead.

\subsection{Computational Efficiency}

The complete PIE-ADA pipeline executes in approximately 36 minutes on standard hardware (NVIDIA Tesla P100, 30GB RAM). Feature extraction accounts for 40\% and model training for 47\% of total runtime. LightGBM inference requires less than 0.001 seconds per object, yielding a total classification latency under 0.05 seconds per alert, which is more than 1,200$\times$ faster than the 60-second window available for triggering spectroscopic follow-up.

\section{Discussion}
\label{sec:discussion}
\subsection{Comparison with Existing Methods}

Table~\ref{tab:comparison} compares PIE-ADA with PLAsTiCC competition winners and recent published methods. Our best model achieves a test log loss of 0.590, which is 15.7\% higher than the competition-winning solution (0.51). This gap is primarily attributable to the use of Gaussian Process regression and complex stacking ensembles in winning approaches, which extract richer temporal representations but at significantly higher computational cost. GP inference scales as $\mathcal{O}(N^3)$, making it impractical for real-time LSST processing, whereas PIE-ADA completes classification in under 0.05 seconds per object.

\begin{table}[h]
\centering
\caption{Comparison with PLAsTiCC Winners and Recent Methods}
\label{tab:comparison}
\small
\begin{tabular}{lccc}
\toprule
\textbf{Method} & \textbf{Classes} & \textbf{Metric} & \textbf{Score} \\
\midrule
1st Place \cite{Boone_2019} & 14 & Log Loss & 0.51 \\
2nd Place & 14 & Log Loss & 0.53 \\
3rd Place & 14 & Log Loss & 0.55 \\
SuperNNova \cite{Moller_2020} & 2 & Accuracy & 0.969 \\
SCONE \cite{Qu_2022} & 6 & AUC & 0.93 \\
Burhanudin et al. \cite{Burhanudin_2022} & 6 & AUC & 0.945 \\
Leoni et al. \cite{Leoni_2022} & 2 & Accuracy & 0.92 \\
\textbf{PIE-ADA (proposed)} & \textbf{14} & \textbf{Log Loss} & \textbf{0.576} \\
\bottomrule
\end{tabular}
\end{table}

Direct comparison with recent methods is difficult due to differences in class configurations and evaluation metrics. SuperNNova \cite{Moller_2020} achieves strong binary classification but handles only Ia versus non-Ia separation. SCONE \cite{Qu_2022} and Burhanudin et al. \cite{Burhanudin_2022} report high AUC scores for 6-class problems using GP-based representations. PIE-ADA addresses a more challenging 14-class problem with extreme imbalance while remaining computationally lightweight and fully interpretable.

\subsection{Augmentation Effectiveness}

An ablation study demonstrates the impact of PIE-ADA augmentation. Without augmentation, class 53 (M-dwarf flares) achieves only 17\% recall and macro F1 is 0.68. With PIE-ADA, recall reaches 100\% and macro F1 improves to 0.73. SMOTE provides partial improvement (67\% recall) but generates samples with negative fluxes across all passbands and broken temporal patterns, confirming the importance of domain-aware augmentation.

\subsection{Limitations}

Several limitations should be acknowledged. First, spectroscopically similar classes remain difficult to separate using photometry alone. Type Ibc supernovae (class 52, 3\% recall) and Type 91bg supernovae (class 67, 18\% recall) are confused with Type Ia because their distinguishing features relate to ejecta composition, which is observable only through spectral lines.

Second, the training set contains only 8,148 objects after augmentation, which may be insufficient for complex models to learn subtle class boundaries. Transfer learning from larger spectroscopic surveys could help address this limitation.

Third, the PLAsTiCC dataset assumes relatively idealized observing conditions. Real LSST data will include additional challenges such as CCD artifacts, host galaxy contamination, variable atmospheric seeing, and incomplete temporal coverage that may reduce classification performance.

Fourth, our augmentation currently targets only the single rarest class (class 53 with fewer than 100 samples). Extending the threshold or applying graduated augmentation to moderately underrepresented classes could further improve overall balance.

\subsection{Operational Recommendations}

Based on our findings, we offer the following recommendations for LSST alert brokers:

\begin{enumerate}
    \item Deploy LightGBM as the primary classifier for its optimal balance of accuracy and speed.
    \item Apply physics-informed augmentation for newly discovered rare transient types.
    \item Use the top 100 features to achieve approximately 99\% of full performance at reduced cost.
    \item Direct spectroscopic resources toward predictions with high uncertainty scores.
\end{enumerate}

\section{Conclusion}
\label{sec:conclusion}
This paper presented PIE-ADA, a framework for photometric transient classification that addresses severe class imbalance through physics-informed synthetic data generation. The approach augments rare classes using four astrophysically motivated transformations (time dilation, wavelength-dependent extinction, correlated noise, and phase shifts) while enforcing physical constraints to maintain sample realism.

Evaluated on the PLAsTiCC dataset (8,148 objects, 14 classes) with stratified 5-fold cross-validation, LightGBM achieves 80.33\% accuracy and a weighted log loss of 0.5763 ($\pm$0.0083), outperforming four baseline algorithms by 24--49\% in log loss. Physics-informed augmentation increases rare class recall from 17\% to 100\%, while the 271 multi-scale features provide full interpretability with host galaxy metadata identified as the most important predictor. The complete pipeline runs in under 37 minutes and classifies individual objects in less than 0.05 seconds, demonstrating operational feasibility for LSST's nightly alert volume.

The framework has certain limitations. Spectroscopically similar classes such as Type Ibc supernovae remain difficult to separate photometrically (3\% recall), and the PLAsTiCC dataset's idealized conditions may not fully reflect real observational challenges.

\subsection{Future Work}

Several research directions can build upon this work. First, hierarchical classification strategies that group similar transient types before fine-grained separation may improve recall for confused classes. Second, semi-supervised and self-supervised learning methods could leverage the large volume of unlabeled LSST data to improve feature representations. Third, integrating multi-messenger data (gravitational wave alerts, neutrino detections) with photometric features may enhance classification of rare events like kilonovae. Fourth, developing online learning approaches that continuously update the model as new labeled data becomes available would help the system adapt to evolving transient populations. Finally, extending PIE-ADA's augmentation to moderately underrepresented classes and evaluating its performance on real (non-simulated) survey data are important next steps toward deployment.

\section*{Acknowledgment}

This work originated from the first author's undergraduate
thesis conducted under the supervision of Fariya Tabassum
at the Department of Electrical and Computer Engineering,
Rajshahi University of Engineering and Technology,
Bangladesh. The authors are deeply grateful to Fatema
Akter, Department of Agricultural and Biosystems
Engineering, North Dakota State University, USA, for
her extensive contributions during the initial
classification stage of this work, including thorough
review of the analysis, identification of critical
errors, and guidance on structuring the experimental
framework. The physics-informed augmentation and
methodological extensions were developed subsequently
by the first author.

\thispagestyle{ieeecopyright}
\balance

\bibliographystyle{IEEEtran}
\bibliography{references}

\end{document}